# Cancer and nonextensive statistics

Jorge A. González [1] and Irving Rondón

*Centro de Física, Instituto Venezolano de Investigaciones Científicas, Apartado Postal 21827, Caracas 1020-A, Venezuela*

**Abstract**

We propose a new model of cancer growth based on nonextensive entropy. The evolution equation depends on the nonextensive parameter $q$. The exponential, the logistic, and the Gompertz growth laws are particular cases of the generalized model. Experimental data of different tumors have been shown to correspond to all these tumor-growth laws. Recently reported studies suggest the existence of tumors that follow a power law behavior. Our model is able to fit also these data for $q < 1$. We show that for $q < 1$, the commonly used constant-intensity therapy is unable to reduce the tumor size to zero. As is the case of the Gompertzian tumors, for $q < 1$ a late-intesification schedule is needed. However, these tumors with $q < 1$ are even harder to cure than the Gompertzian ones. While for a Gompertzian tumor a linearly-increasing cell-kill function is enough to reduce the tumor size to zero following an exponential decay, in the case of tumors with $q < 1$, the exponential decay is obtained only with an exponentially increasing cell-kill function. This means that these tumors would need an even more aggressive treatment schedule. We have shown that for Gompertzian tumors a logarithmic late-intensification is sufficient for the asymptotic reduction of the tumor-size to zero. This is not the fastest way but it is more tolerable for patients. However for the tumors with $q < 1$ we would need at least a linearly increasing therapy in order to achieve a similarly effective reduction. When $q > 1$, tumor size can be reduced to zero using a traditional constant-intensity therapy.

## 1 Introduction

Many models of tumor growth have been proposed to fit experiments and clinical data [1]-[13]. Despite considerable progress in understanding tumor development, the law of growth for human tumors is still a matter of debate

[1] Corresponding author. Fax: +58-212-5041148; e-mail: jorge@ivic.ve

[13]- [16]. Some of the most important tumor growth models are the exponential, logistic and Gompertz laws [1]-[16]. Very recently, Hart et al [13] have studied clinical data of some tumors that support power law growth. The most popular of all tumor growth models is the Gompertzian law [1]-[5,12]. In the famous paper [1], Norton et al say that the growth of most experimental neoplasms and human tumors are well described by the Gompertz equation. However, as mentioned earlier, there are many works presenting different types of tumors that challenge the Gompertz model [13]

Despite the immense use of the Gompertzian law in scientific research, there has been always the problem of deriving it from theoretical considerations based on the bio-physics of the growth [17]. In Ref. [17] a theoretical justification for the Gompertz's law of the growth was presented. This justification is based on the concept of entropy associated with malignat growth. The entropy formula used in Ref. [17] is the well-known Boltzmann- Gibbs extensive entropy. Recently, there has been a wealth of work dedicated to nonextensive statistics [18]-[23]. The entropy introduced in Ref. [18] gives rise to new and interesting effects (which would be relevant for the description of thermodynamically anomalous systems). There are already many successful applications of this new theory in physics [20]-[25]

We will use the new nonextensive entropy in the derivation of a new very general evolution equation for tumors. We will show that the logistic, Gompertz, exponential and power laws are particular cases of the new equation. Everything depends on the nonextensive parameter $q$ [18]-[25]. This suggests that different types of tumors may possess different values of the nonextensive parameter $q$. In Ref [26] De Vladar and González have investigated the dynamical response of Gompertzian tumors under the influence of immunological activity and therapy. They have shown that immunological activity alone is not sufficient to induce a complete regression of the tumor. Very small tumors always tend to grow. Since immunity cannot induce a complete tumor regression, a therapy is always required. However only late-intensification therapies can be succesful.

We will investigate the obtained nonlinear equation under the action of different treatments. We will show that logarithmic late intensification can reduce a Gompertzian tumor cell population to zero asymptotically. We will prove that the tumors for which $q < 1$ need a much more agressive treatment schedule than the Gompertzian ones. The recently reported tumors by Hart et al [13] in an experimental study correspond to a nonextensive parameter $q < 1$.

## 2 Tumor growth and Boltzmann - Gibbs entropy

In this section we will present a brief summary of the ideas of the work [17]. Tumor growth is restricted by the tissue's carrying capacity. At the beginning the growth is fast. Then, in the long time behavior, the cell population saturates: the cell population reaches a maximum asymptotically. This maximum is determined by the tissue's carrying capacity. The dynamics of a tumor should depend on the relation between active and resting cells. Suppose $P_1$ is the probability for a tumor cell to be active, and $P_2$ is the probability for a tumor cell to be at rest in such way that $P_1 + P_2 = 1$. There is experimental evidence that the proportion of resting cells increases as the growth of the tumor progresses.

In this case, the Boltzmann - Gibbs entropy is defined as:

$$S = -\kappa_b \left( P_1 \ln P_1 + P_2 \ln P_2 \right), \tag{1}$$

where $\kappa_b$ is certain constant.

The authors of Ref [17] postulate that the rate of change of $P_2$ is proportional to the entropy:

$$\frac{dP_2(t)}{dt} = CS(t), \tag{2}$$

where $C$ is certain constant.

Suppose that $N(t)$ is the population of the resting tumor cells. In Ref. [17] the authors put

$$P_2(t) \approx \frac{N(t)}{N_\infty}, \tag{3}$$

where $N_\infty$ is the asymptotic value of $N(t)$ when $t \to \infty$.

From equation (2), the authors of Ref. [17] obtained an approximation to the Gompertz equation :

$$\frac{dN}{dt} = -\kappa N \ln \left( \frac{N}{N_\infty} \right), \tag{4}$$

where $\kappa = \kappa_b C$.

## 3 Tsallis statistics

Boltzmann - Gibbs statistics satisfactorily describes nature if the microscopic interactions are short-ranged and the effective microscopic memory is short-ranged and the boundary conditions are nonfractal. There is a large series of recently found physical systems that present anomalies that violate the standard Boltzmann- Gibbs method. A nonextensive thermostatistics, which contains the Boltzmann- Gibbs as a particular case was proposed and developed by Tsallis and coworkers in a series of papers [18]-[23]. Nowadays, physicists have produced a large amount of successful applications of Tsallis statistics [20]-[25]. This includes systems with long-range interactions, long-range microscopic memory, and systems which possess a fractal or multifractal structure [20]-[25].

The Tsallis entropy is defined as follows :

$$S_q = \kappa \frac{1 - \sum_{i=1}^{w} P_i^q}{q - 1}, \qquad (5)$$

where $\kappa$ is a positive constant, $w$ is the total number of possibilities of the system, $\sum_{i=1}^{w} P_i = 1$, $q \ \epsilon \ R$. This expression recovers the Boltzmann-Gibbs entropy, $S_1 = -\kappa \sum_{i=1}^{w} P_i \ln P_i$ in the limit $q \to 1$. Parameter $q$ characterizes the degree of nonextensivity of the system.

This can be seen in the following rule :

$$S_q(A+B)/\kappa = [S_q(A)/\kappa] + [S_q(B)/\kappa] + (1-q)[S_q(A)/\kappa][S_q(B)/\kappa], \qquad (6)$$

where $A$ and $B$ are two independent systems in the sense that $P_{ij}(A+B) = P_i(A)P_j(B)$.

## 4 A new generalized evolution equation for tumor growth

Considering the same ideas discussed in Section 2 for the justification of Gompertz's equation introduced by Calderon and Kwembe in Ref. [17], but using the nonextensive entropy defined in Eq. (5), we obtain the following new evolution equation for tumor growth :

$$\frac{dN}{dt} = \frac{\kappa N_\infty}{q-1} \left[ 1 - \left(\frac{N}{N_\infty}\right)^q - \left(1 - \frac{N}{N_\infty}\right)^q \right]. \qquad (7)$$

Note that if $q = 2$, Eq (7) is equivalent to the logistic equation: $\frac{dN}{dt} = 2\kappa \left[ N - \frac{N^2}{N_\infty} \right]$. In the limit $N_\infty \to \infty$, we get the exponential law: $\frac{dN}{dt} = 2\kappa N$. In the limit $q \to 1$, from Eq (7) we obtain the following equation:

$$\frac{dN}{dt} = -\kappa \left[ N \ln \left( \frac{N}{N_\infty} \right) + (N_\infty - N) \ln \left( 1 - \frac{N}{N_\infty} \right) \right]. \tag{8}$$

This equation has been considered as an aproximation to the Gompertz's equation for large values of $N(t)$ [17]. However, we should say that some of the most important consequences of the Gompertz's law used in the design of new treatments are obtained from the behavior of the Gompertz's equation near the point $N = 0$. And Eq. (8) possesses exactly the same property as the Gompertz's equation when $N = 0$. Morever, Eq. (8) is as good as Gompertz's equation in the overall fitting of the available experimental data.

Another important case of the equation (7) is when $0 < q < 1$. For $\frac{N}{N_\infty} \ll 1$, equation (8) yields:

$$\frac{dN}{dt} = \left( \frac{\kappa N_\infty}{1 - q} \right) \left( \frac{N}{N_\infty} \right)^q. \tag{9}$$

Equations (7) and (9) imply a power law growth for small and middle-sized tumors. And this is exactly what is obtained from the clinical data investigated by Hart et al in Ref. [13]. Figures 1 show the behavior of the solution to equation (7) for different values of the parameters. A "pure" power law $\frac{dN}{dt} = \kappa N^\beta$ for all the time would imply an unbounded growth, which is impossible from the biophysical point of view. So Equation (7) is a very nice solution to this modelling problem, because for $0 < q < 1$ it yields a power law growth, but for $t \to \infty$, the tumor size will be bounded as expected (for instance, due to finite tissue's carrying capacity).

## 5  Cancer treatment schedules

In the treatment of avanced tumors, many therapy schedules employ intensive therapy initially, when the tumor is very large, then the dose is decreased as tumor is reduced. In other treatments, the dose is maintained constant during the therapy course. In the post-surgical setting, where it is thought that only microscopic loci of tumor are left, the dose schedule of adjuvant chemotherapy chosen is often less intense than would be used for a larger tumor of equivalent type. Many observations strongly suggest that the microscopic loci left after surgical resection are sufficiently resistant to chemotherapy and so they are not easily curable by low exposure. Thus, some authors [6][10]-[12][16] have suggested the use of more intense schedules, higher doses, and more prolonged therapy than the normally used.

The level of therapy adequate to initiate regression may not be sufficient to sustain regression and produce cure. The result is that many cancers are impossible to cure using the common treatments [6][10]-[12][16]. Norton and coworkers [6][10,11] have suggested that one way to combating the slowing rate of regression of a tumor as it shrinks in response to therapy was to increase the intensity of treatment as the tumor became smaller. This kind of treatment schedule is called " late intensification". It can be thought that the smallest resulting tumor size for a given total dose of therapy is accomplished if the entire therapy is given over as short a time period as possible. Because of toxicity, however, this is generally impossible [6].

Suppose that the growth of an untreated solid tumor is described by the equation $\frac{dN}{dt} = f(N)$. We now proceed by adding a cell-kill term to this equation to represent the effect of treatment :

$$\frac{dN}{dt} = f(N) - C(t)N, \tag{10}$$

where $C(t)$ is proportional to the drug concentration in chemotherapy and to radiation dose in radiotherapy [6].

In the case of the Gompertz equation ( $f(N) = -\kappa N \ln \frac{N}{N_\infty}$) and with constant drug concentration throughout the time of interest ($C(t) = C_0$), there is a very simple solution

$$N(t) = N_0 \exp\left[\left(\ln \frac{N_\infty}{N_0} - \frac{C_0}{\kappa}\right)\left(1 - e^{-\kappa t}\right)\right]. \tag{11}$$

In the untreated case, it is well known that $N_\infty$ is the asymptotic limit for $N(t)$ when $t \to \infty$. When $C_0 \neq 0$, there is a new limit $N(t) \to N_\infty e^{-\frac{C_0}{\kappa}}$. For finite $C_0$, this value is always larger than zero. Figure 2 a) shows that a constant $C(t)$ is unable to reduce $N(t)$ to zero. Note that in this case and in all the other simulations where the evolution equation contains a cell-kill term, we have allowed the tumor to grow untreated for and interval of time and then we have applied the treatment. Wheldon [6] has proposed a linear increasing function of time for $C(t) = A + Bt$, where $A = \omega - \kappa \ln \frac{N_\infty}{N_0}$, $B = \kappa\omega$ and $\omega$ is some positive constant. With this increasing treatment, tumor depopulation will proceed as:$N(t) = N_0 e^{-\omega t}$. However, a linear increasing function for dose intensity can be very difficult for the patients. Figure 2 b) shows that $N(t)$ tends to zero when $C(t) = A + Bt$.

# 6 Logarithmic cell-kill function

Can we find an intesification schedule able to reduce to zero asymptotically the tumor cell population, but using a slowly increasing function of time for $C(t)$ ?. Suppose we have the following therapy: $C(t) = C_0 \ln(e + \delta t)$. This problem has been addressed in Ref. [27]. In general, it is possible to prove that for a therapy of this type, the population will tend to zero as a potential decay.

This can be shown using the general solution to equation

$$\frac{dN}{dt} = -\kappa N \ln\left(\frac{N}{N_\infty}\right) - C(t)N, \tag{12}$$

which can be written in the following way

$$N(t) = N_\infty \exp\left[e^{-\kappa t}\left(\ln\frac{N}{N_\infty} - \int_0^t e^{\kappa s}C(s)ds\right)\right]. \tag{13}$$

When $C(t)$ is defined as the logarithmic function, we have

$$N(t) = N_\infty \exp\left[e^{-\kappa t}\left(\ln\frac{N}{N_\infty} - \int_0^t e^{\kappa s}C_0\ln(e + \delta S)ds\right)\right]. \tag{14}$$

In the asymptotic limit ($t \to \infty$), we obtain $N(t) \simeq N_\infty \exp\left[-\frac{C_0}{\kappa}\ln(\delta t)\right]$, and finally, $N(t) \simeq \frac{N_0}{At^\gamma}$, where $\gamma = \frac{C_0}{\kappa}$, $A = \delta^\gamma$.

In figures 2 c) and 2 d) it is shown that a logarithmic cell-kill function can reduce the tumor size to zero asymptotically. Using the same total amount of therapy ( that is, the same total area under the graph of $C(t)$ ), the schedule following the logarithmic function produces a larger reduction of tumor cell population than the schedule using a constant intensity therapy. Besides, it is expected that the logarithmic intesification should be more tolerable for the patients than a linear increasing- function-of-time therapy. It is important to remark that the rate of increment of the logarithmic function decreses with time. Moreover, this rate does not depend on parameter $\delta$.

Althought late intensification schedules have been used in real trials [28]-[33] with many patients and with great success in general, the actual employed schedules are similar to a step function with only two values for the dose intensity: a normal dose and a higher dose. Except for this step, the dose is almost always mantained constant. On the other hand, it is not uncommon to find treatment schedules where the dose is decreased as tumor is reduced. Our analysis shows that it is very important to intensify therapy as much as posible, and as long as possible.

Using the general solution to equation (12)

$$N(t) = N_\infty \exp\left[e^{-\kappa t}\left(\ln\frac{N}{N_\infty} - \int_0^t e^{\kappa s}C(s)ds\right)\right]$$

and the same relationship that represent equation (12).i.e $\frac{dN}{dt} = -\kappa N \ln\left(\frac{N}{N_\infty}\right) - C(t)N$ we can prove that, to induce tumor regression monotonically (that is $\frac{dN}{dt} < 0$), the treatment must be increased continuously. Furthermore the only way to attain an asymptotic decrease of tumor size to zero is through an intesified therapy.

## 7 Tumors with $0 < q < 1$

Now let us consider the following general equation with a cell-kill term

$$\frac{dN}{dt} = \frac{\kappa N_\infty}{q-1}\left[1 - \left(\frac{N}{N_\infty}\right)^q - \left(1 - \frac{N}{N_\infty}\right)^q\right] - C(t)N. \tag{15}$$

Following the same ideas discussed in Section 6, we can prove that tumors with $0 < q < 1$ require also a time-increasing therapy in order to reduce $N(t)$ to zero. In fact, $C(t)$ must behave as

$$C(t) = \frac{\kappa N_\infty^{1-q}}{1-q}\frac{1}{N^{1-q}}, \tag{16}$$

as $N \to 0$ in order to sustain tumor reduction. But there are other more disturbing facts in the behavior of Eq. (15). Figure 3 a) shows that a constant cell-kill function cannot reduce to zero the tumor size $N(t)$ in the framework of Eq. (15) with $q = 0.5$.

If we require $N(t)$ to be an exponentially decreasing function, then we obtain that $C(t)$ must increase exponentially. This is very unlike the Gompertzian tumors, where a linear increasing theraphy was sufficient for an exponentially decreasing tumor. In figure 3 b) the solution to equation (15) with $C(t) = C_0 \exp(\frac{\omega}{2}t)$ is shown. In this case $N(t)$ behaves as an exponentially decreasing function. On the other hand, for a tumor to be reduced following a power law, say: $N \sim \frac{N_0}{\alpha t^\gamma}$, we would need an increasing power-law therapy $C(t) \sim t^{\gamma(1-q)}$. Note that if $q = 1/2$ and $\gamma = 2$, the needed therapy will be a linearly increasing function of time. For a faster decreasing tumor (say $\gamma = 4$), the required therapy should increase as a quadratic function.

Figure 3 c) shows different solutions to Eq. (15) for different cell-kill functions of type $C(t) = C_0 [1 + \delta t]^{\gamma(1-q)}$. Note that this kind of treatment can be successful in some cases even for $q = 0.5$. In particular, the linear and quadratic cell-kill functions can reduce the tumor size at reasonable rates. However, the logarithmic cell-kill function (that was used before in the Gompertzian tumors) is not successful for the tumors with $q = 0.5$. This can be observed in Figure 3 d). All this means that the tumors with $0 < q < 1$ can be very hard to cure.

Let us investigate the equation $\frac{dN}{dt} = \alpha N^{1/2}$, that was reconstructed from their data by Hart et al [13]. Note that this equation is a particular case of equation (9). The exact solution this equation is $N(t) = \left(\sqrt{N_0} + \frac{\alpha}{2}t\right)^2$, where $N_0$ is the initial condition for $t = 0$.

If we consider a cell - kill term as in Eq. (15), we have

$$\frac{dN}{dt} = \alpha N^{1/2} - C(t)N. \tag{17}$$

Using the transformation $N = M^2$, we get a linear equation $\frac{dM}{dt} = \frac{\alpha}{2} - C(t)M$. So the general solution to the equation (17) is

$$N(t) = M^2(t), \tag{18}$$

where $M(t) = e^{-\int_0^t C(s)ds} \left[\int_0^t \frac{\alpha}{2} e^{\int_0^S C(r)dr} dS + N_0^{1/2}\right]$.

For a constant therapy $C(t) \equiv C_0$, we have $M = \left(N_0^{1/2} - \frac{\alpha}{2C_0}\right) e^{-C_0 t} + \frac{\alpha}{2C_0}$.

Note that for $t \to \infty$, $N(t) \to \frac{\alpha^2}{4C_0^2}$. It is easy to check that if we require $N(t) \to 0$ in Eq.(17), we need a continuously increasing function for $C(t)$. If we require a tumor - size decay as $N(t) \sim e^{-\omega t}$, the function $C(t)$ must increase as $C(t) \sim e^{\frac{\omega}{2}t}$. On the other hand, a linearly increasing $C(t)$ would lead only to a power - law decay of the tumor size. Additionally, any decreasing $C(t)$ would lead to an increasing tumor size. Even a slowly decreasing $C(t)$ as $C(t) \sim \frac{1}{t}$ would lead to a tumor size increasing as $N(t) \sim t$.

Finally, a logarithmic intensification of the dose intensity, e.g. $C(t) = C_0 \ln(e + \delta t)$, produces a tumor - size decay that behaves asymptotically as $N(t) \sim \frac{1}{4C_0^2 \ln^2(\delta t)}$, which is too slow to be effective. So, perhaps for this kind of tumor, the most reasonable schedule is one that is a linearly - increasing cell - kill function. Such a schedule is not so agressive as the exponential one. Meanwhile it achieves an asymptotic reduction of the tumor-size to zero following a power-law decay. The behavior of the solution to Eq. (17) for different cell-kill functions $C(t)$ can be observed in Figure 4.

## 8 Tumors with $q > 1$

Investigating equation (15) when $q > 1$, we obtain the result that the tumor size can be reduced to zero using a constant-intensity therapy $C(t) = C_0$, provided that $C_0 > \frac{\kappa q}{q-1}$. Thus, in comparision with the tumors with $q \leq 1$, the tumors with $q > 1$ are relatively easy to cure. An example is shown in the Fig. 5

## 9 Conclusions

Based on ideas related to nonextensive entropy, we have obtained a new evolution equation for tumor size. This evolution equation depends on the non extensive parameter $q$. For different values of $q$, the generalized evolution equation reduces to Gompertz-like , logistic, exponential and power law equations. Experimental data of different tumors have been shown to correspond to all these tumor-growth laws. We believe there are different types of tumors in nature and the new evolution equation can help to fit the existing and future experimental data in order to make important predictions in medical practice and to design new treatments.

This is specially important since, recently, tumors increasing following a power-law have been reported. These tumors seem to contradict the more traditional exponential and Gompertz models. The success of the generalized equation in fitting the new data could be an indication that these tumors possess some fractal behavior because the nonextensive entropy is often related to fractal systems. We have investigated the dynamics of the nonlinear equation under the action of different cell - kill functions .We have shown that logarithmic late intensification can reduce a Gompertzian tumor cell population to zero asymptotically. This is very relevant to clinical practice because the usual constant-function therapy is unable to reduce to zero the tumor size [10,12][26]. The size of tumors that follow equation (15) with $q > 1$ can be reduced to zero using a traditional constant-intensity therapy.

However, it is very disturbing that the tumors with $q < 1$ are very hard to cure. While for a Gompertzian tumor a linearly - increasing cell-kill function is enough to reduce the tumor - size to zero following an exponential decay;

in the case of the tumors with $q < 1$, the exponential decay is obtained only with an exponentially increasing cell-kill function. A linearly increasing cell - kill function would lead to a power-law decay. All this could explain why even the high-dose late-intensification schedules applied following Norton studies [28]-[33] have been unsuccessful in many cases [33]. These tumors would need an even more aggressive treatment schedule.

List of figures

(1) Behavior of the solution to equation (7) for different parameter values a) $N_\infty = 15$, $\kappa = 1$ are fixed; $q = 0.5$ for I , $q = 1.1$ for II , $q = 2$ for III, b) $q = 2$, $\kappa = 1$ are fixed; $N_\infty = 100$ for I , $N_\infty = 20$ for II, $N_\infty = 15$ for III ; c) $q = 0.5$ and $\kappa = 1$ are fixed; $N_\infty = 100$, for I, $N_\infty = 20$ for II, $N_\infty = 15$ for III; d) $q = 0.5$ and $N_\infty = 15$ are fixed; $\kappa = 10$ for $I$, $\kappa = 1$ for II, $\kappa = 0.1$ for III.
(2) The solution to the Gompertz equation $N(t)$ (Eq. (10) with $f(N) = -\kappa N \ln \frac{N}{N_\infty}$) a The asymptotic limit of $N(t)$ with a constant $C(t)$ is always finite. b) The solution $N(t)$ tends to zero when $C(t) = A + Bt$. c) A Gompertzian tumor can be reduced to zero asymptotically using a logarithmic cell-kill function $C(t) = C_0 \ln(e + \delta t)$, $C_0 = 1.5$, $\delta = 0.5$. d) The same as in c), with different parameters $C_0 = 2$, $\delta = 2$.
(3) Behavior of the solution to Eq.(15) for different cell-kill functions with $q = 0.5$. a) A constant cell-kill function cannot reduce to zero the tumor size $N(t)$ if $q = 0.5$. b) The solution can be an exponentially decreasing function if $C(t) = C_0 \exp(\frac{\omega}{2} t)$. c) The solution can tend to zero asypmptotically if $C(t) = C_0 [1 + \delta t]^{\gamma(1-q)}$; I ) $\gamma = 1$, II ) $\gamma = 2$, III ) $\gamma = 4$. d) A logarithmic cell-kill function is not successful for a tumor described by Eq. (15) with $q = 0.5$.
(4) Different tumor evolutions in the framework of Eq. (17) with cell-kill functions a) $C(t) = C_0$; b) $C(t) = C_0 \exp(\frac{\omega}{2} t)$; c) $C(t) = C_0 \ln(e + \delta t)$; d) $C(t) = C_0 [1 + \delta t]^{\gamma/2}$.
(5) The solution to Eq. (15) when $q > 1$, $C(t) = C_0 = const.$ a) $C_0 < \frac{\kappa q}{q-1}$, $q = 2$ for I ; $q = 2.5$ for II ; b) $C_0 > \frac{\kappa q}{q-1}$, $q = 2$ for I ; $q = 2.5$ for II . Note that, in this case, the tumor-size can be reduced to zero.

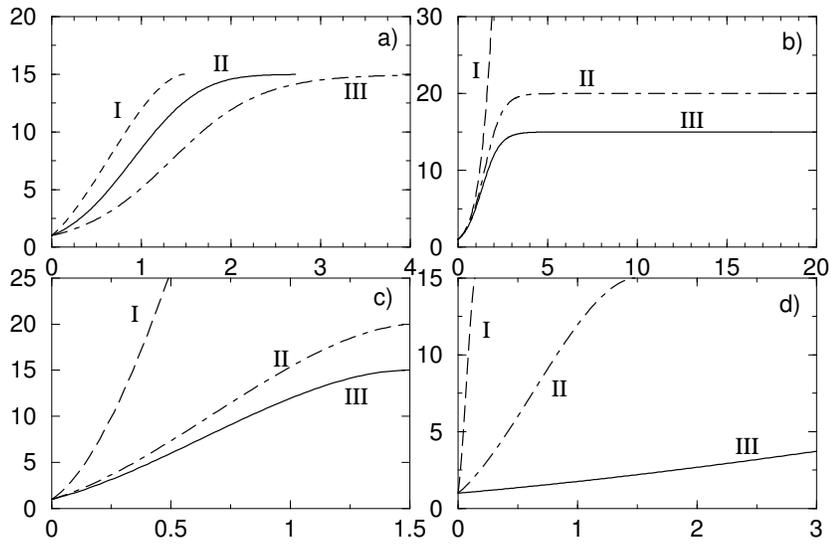

Fig. 1. Behavior of the solution to equation (7) for different parameter values a) $N_\infty = 15$, $\kappa = 1$ are fixed; $q = 0.5$ for I , $q = 1.1$ for II , $q = 2$ for III, b) $q = 2$, $\kappa = 1$ are fixed; $N_\infty = 100$ for I , $N_\infty = 20$ for II, $N_\infty = 15$ for III ; c) $q = 0.5$ and $\kappa = 1$ are fixed; $N_\infty = 100$, for I, $N_\infty = 20$ for II, $N_\infty = 15$ for III; d) $q = 0.5$ and $N_\infty = 15$ are fixed; $\kappa = 10$ for I, $\kappa = 1$ for II, $\kappa = 0.1$ for III.

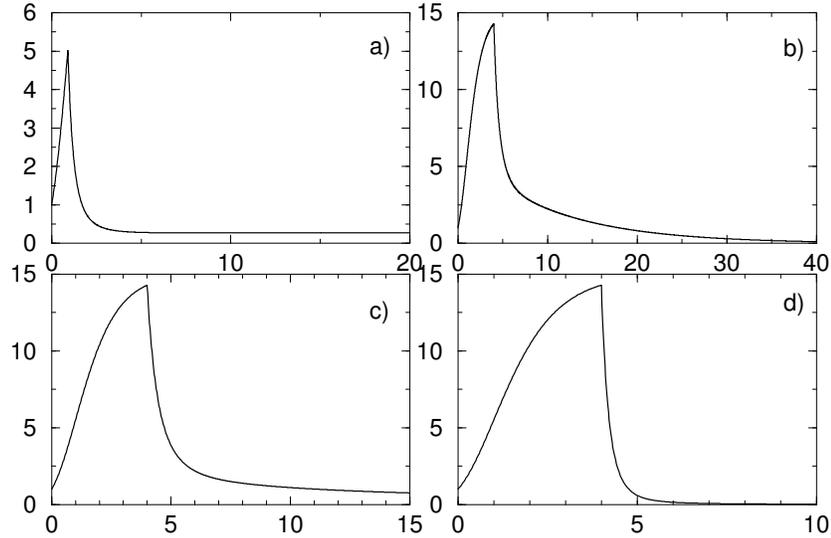

Fig. 2. The solution to the Gompertz equation $N(t)$ (Eq. (10) with $f(N) = -\kappa N \ln \frac{N}{N_\infty}$) a The asymptotic limit of $N(t)$ with a constant $C(t)$ is always finite. b) The solution $N(t)$ tends to zero when $C(t) = A + Bt$. c) A Gompertzian tumor can be reduced to zero asymptotically using a logarithmic cell-kill function $C(t) = C_0 \ln(e + \delta t)$, $C_0 = 1.5$, $\delta = 0.5$. d) The same as in c), with different parameters $C_0 = 2$, $\delta = 2$.

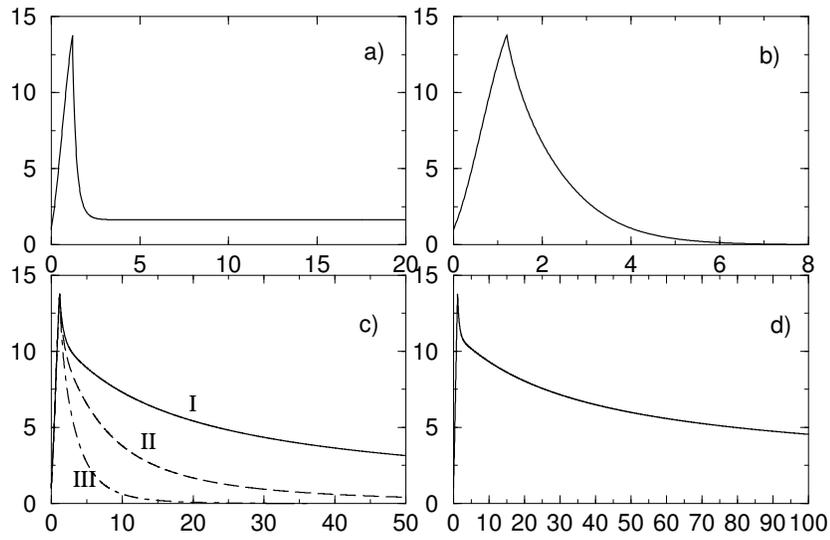

Fig. 3. Behavior of the solution to Eq.(15) for different cell-kill functions with $q = 0.5$. a) A constant cell-kill function cannot reduce to zero the tumor size $N(t)$ if $q = 0.5$. b) The solution can be an exponentially decreasing function if $C(t) = C_0 \exp(\frac{\omega}{2}t)$. c) The solution can tend to zero asypmptotically if $C(t) = C_0[1 + \delta t]^{\gamma(1-q)}$; I ) $\gamma = 1$, II ) $\gamma = 2$, III ) $\gamma = 4$. d) A logarithmic cell-kill function is not successful for a tumor described by Eq. (15) with $q = 0.5$.

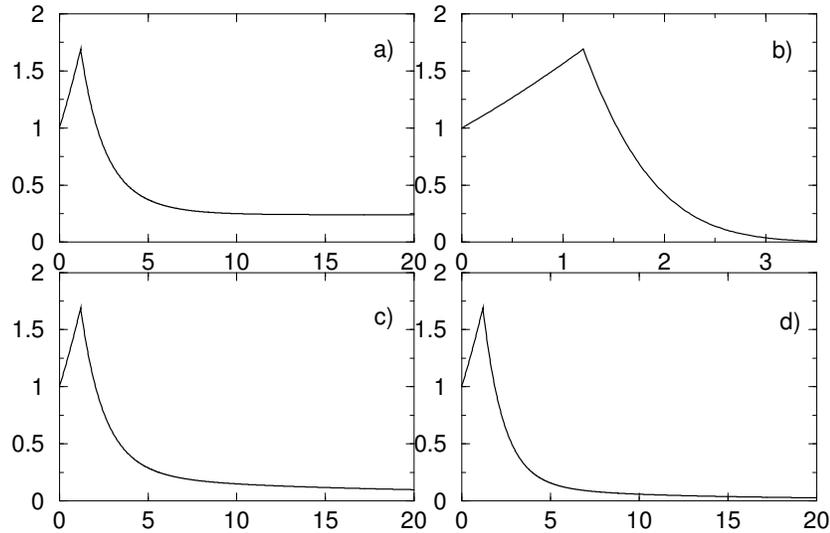

Fig. 4. Different tumor evolutions in the framework of Eq. (17) with cell-kill functions a) $C(t) = C_0$; b) $C(t) = C_0 \exp(\frac{\omega}{2}t)$; c) $C(t) = C_0 \ln(e + \delta t)$; d) $C(t) = C_0[1 + \delta t]^{\gamma/2}$.

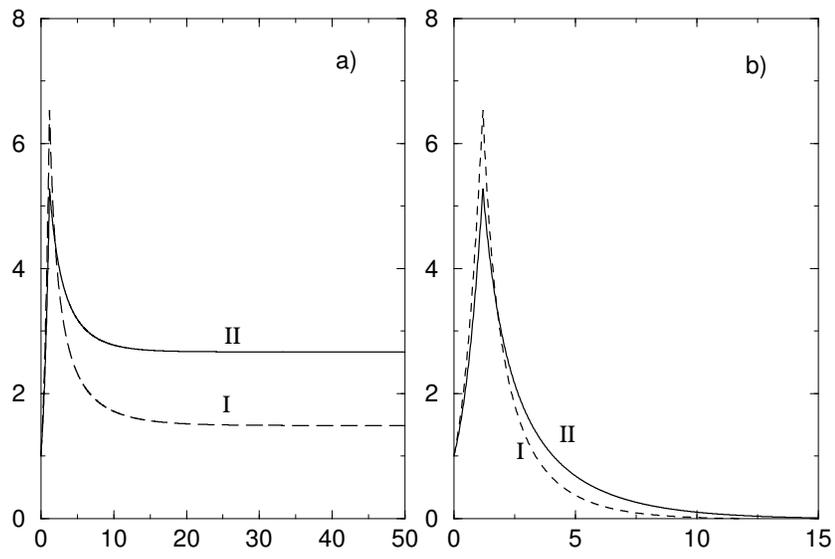

Fig. 5. The solution to Eq. (15) when $q > 1$, $C(t) = C_0 = const.$ a) $C_0 < \frac{\kappa q}{q-1}$, $q = 2$ for I ; $q = 2.5$ for II ; b) $C_0 > \frac{\kappa q}{q-1}$, $q = 2$ for I ; $q = 2.5$ for II . Note that, in this case, the tumor-size can be reduced to zero.